\documentclass{aa}
\usepackage[varg]{txfonts}
\usepackage[utf8]{inputenc}\usepackage[T1]{fontenc}
\usepackage{graphicx}
\usepackage{hyperref}

\newcommand{\msun}{\ensuremath{M_{\odot}}}

\def\apjl{{\rm ApJ}}                
\def\apjs{{\rm ApJS}}               
\def\ao{{\rm Appl.~Opt.}}           
\def\aap{{\rm A\&A}}                

\begin{document}

\title{Image segmentation for analyzing galaxy-galaxy\\strong lensing systems}

\author{Bryan Ostdiek
\and Ana Diaz Rivero
\and Cora Dvorkin}
\institute{
Department of Physics, Harvard University, Cambridge, MA 02138, USA
}
\titlerunning{Image segmentation for analyzing galaxy-galaxy strong lensing systems}
\authorrunning{Ostdiek et al.}

\date{\today}
        
\abstract{}
{
The goal of this paper is to develop a machine learning model to analyze the main gravitational lens and detect dark substructure (subhalos) within simulated images of strongly lensed galaxies.
}
{Using the technique of image segmentation, we turn the task of identifying subhalos into a classification problem, where we label each pixel in an image as coming from the main lens, a subhalo within a binned mass range, or neither.
Our network is only trained on images with a single smooth lens and either zero or one subhalo near the Einstein ring.
}
{On an independent test set with lenses with large ellipticities, quadrupole and octopole moments, and for source apparent magnitudes between 17-25, the area of the main lens is recovered accurately.
On average, only 1.3\% of the true area is missed and 1.2\% of the true area is added to another part of the lens.
In addition, subhalos as light as $10^{8.5}\msun$ can be detected if they lie in bright pixels along the Einstein ring.
Furthermore, the model is able to generalize to new contexts it has not been trained on, such as locating multiple subhalos with varying masses or more than one large smooth lens.
}
{}

\maketitle

\section{Introduction}

In this Letter we introduce a new approach to the analysis of galaxy-galaxy strong lensing systems.
We train a network to do image segmentation, defined as labeling every pixel within an image, to find the outline of the main lens and subhalos.
Strong gravitational lensing occurs when light from a distant source (for example, a galaxy or quasar) is distorted and magnified due to the gravitational influence of a large foreground mass (typically a galaxy or a cluster), which acts as a lens.
The dark matter halo surrounding the foreground galaxy accounts for most of the mass of the lens, and smaller dark matter structures within the lens can cause small perturbations to it.\footnote{Technically, perturbations can also be caused by field halos along the line of sight \citep{D_Aloisio_2010,Li:2016afu,McCully:2016yfe,Despali:2017ksx,Sengul:2020yya}, and while here our main concern is with substructure, our method is fully generalizable to images with line-of-sight perturbers.}
Galaxy-galaxy lenses are particularly interesting because these perturbations are caused by very low-mass halos, which, if resolved, can give us a window into the smallest scales of structure formation and test so-far untested predictions of the standard Lambda cold dark matter (CDM) paradigm.

Traditional techniques to directly find substructure from strong lens images rely on modeling the smooth component of the lensing galaxy and reconstructing the source, subsequently ray-tracing the source through the gravitational potential of the smooth lens and looking for residuals between the generated and observed images that could be explained by local overdensities~\citep{1998MNRAS.295..587M, 2003MNRAS.339..607M, 2005MNRAS.363.1136K, 2009MNRAS.392..945V, 2013ApJ...767....9H}.
Two systems with evidence for substructure have been found using one such direct detection method, gravitational imaging, one with a mass of $\left(3.51 \pm 0.15 \right)\times10^9 \msun$~\citep{2010MNRAS.408.1969V} and one with a mass of $\left(1.9\pm 0.1\right)\times10^8\msun$~\citep{2012Natur.481..341V}, although in theory this method should be sensitive to masses as low as $10^7 \msun$ for a subhalo on the Einstein ring~\citep{2009MNRAS.392..945V}. 
The downside to this method is that the smooth component and source have to be modeled and/or reconstructed, and inaccuracies in the lens model can lead to extra residuals and false positives~\citep{2019MNRAS.485.2179R}. 
The modeling pipeline is time-consuming, and alternative modeling approaches can lead to significantly different results for the same system~\citep{2014MNRAS.442.2017V}.

Despite the difficulty of characterizing the substructure content of a lensing galaxy, the fact that the data come in the form of images makes this problem an ideal candidate for cutting-edge machine learning (ML) techniques that have excelled at image recognition tasks across many domains.
\cite{2020arXiv200812731A} use an unsupervised learning technique to detect subhalos in simulated images. 
In \cite{2020PhRvD.101b3515D}, a convolutional neural network (CNN) was used for binary classification to determine if a strong lens image contained substructure beyond the main lens or not.
\cite{2019ApJ...886...49B} and \cite{2020arXiv200505353V} built networks that assume the presence of a population of substructure and infer its properties.
Finally, \cite{2020ApJ...893...15A} used CNNs to classify and distinguish between different types of dark matter substructure. 

With image segmentation, we can analyze the lens and find substructure within a fraction of a second directly from the image instead of minimizing residuals from the modeled system.\footnote{This requires training the model across a wide variety of lensing scenarios.}
In \cite{Ostdiek:2020mvo}, we utilize this new methodology to count the number of pixels predicted for each subhalo mass in order to determine the subhalo mass function slope for simulated data.

\section{Data simulation and network training}
\label{sec:Accuracy}

Traditional methods for studying strong lensing systems involve modeling the mass profile of the lens, light from the lens, and light from the source.
As a proof-of-concept, we do not include light from the lensing galaxy in our simulations.\footnote{The background galaxy at higher redshift often emits at a different spectrum, allowing one to remove light from the lensing galaxy with less modeling.}

Our lensing images were generated using \texttt{Lenstronomy 1.9.1}~\citep{2018PDU....22..189B,2015ApJ...813..102B}.
We fixed the redshift of the lensing galaxy to $z_{\rm{lens}}=0.2$ and the redshift of the source light to $z_{\rm{source}}=0.6$.
We used images that are $80\times 80$ pixels and cover an area of $5\arcsec \times 5\arcsec$, giving a resolution of $0.06\arcsec$.
Here we describe the main features of our data sets. Further details are found in the appendix.

We modeled the light as an elliptical Sersic profile. 
The ratio of the minor-to-major axis ($q$) was chosen to be between 1/3 and 1, and the angle of the major axis was chosen randomly.

The main lens for our ``simple lens'' systems was modeled with a singular isothermal ellipse (SIE) mass profile~\citep{1994A&A...284..285K}.
The Einstein radius was chosen to be between 0.85\arcsec and 1.5\arcsec, and the ellipticity $q$ was between 0.4 and 1.0.

We also examined more realistic ``complex lenses'' that allow for departures from ellipticity by adding $m=3$ and $m=4$ multipole moments to the lensing potential using the parametrization of~\cite{2013arXiv1307.4220X}.
The complex lenses also contain external shear.

In addition to the smooth lens, we can also include substructure.
We modeled subhalos with a truncated Navarro-Frenk-White (NFW) profile~\citep{1996ApJ...462..563N,2009JCAP...01..015B} with a fixed concentration $c=15$.
We studied subhalos with masses between $10^{5.75}\msun$ and $10^{10.24}\msun$.
When training the U-Net, we placed the subhalos in bright pixels, defined as pixels that are at least $50\%$ as bright as the brightest pixel.
We find that the network trained in this way is able to detect massive subhalos in dimmer locations despite never seeing examples while training.

Our simulated images were made by ray tracing the light from the source through the lens to the observer.
Each simulated image contains a unique lens and source.
We used the \texttt{SimulationAPI} of \texttt{Lenstronomy}, selecting the Hubble Space Telescope (HST) module.
We chose the WFC3\_F160W camera band and used the built-in pixel-based point spread function (PSF).
Two exposure times were studied: 1 orbit and 50 orbits.
We assumed a 90 minute orbit, or 5400 s of exposure for 1 orbit and 270000 s for 50 orbits,  which is simplistic and does not take into account the overhead time required for making the measurements, which can take nearly half an orbit~\citep{dressel2010wide}.
We assumed the sky brightness to be magnitude 22.3 and the zero point magnitude to be 25.96~\citep{2011ApJS..193...27W}.\footnote{These are the default values given by \texttt{Lenstronomy}; see \url{https://docs.google.com/spreadsheets/d/1pMUB_OOZWwXON2dd5oP8PekhCT5MBBZJO1HV7IMZg4Y/edit\#gid=580783795}}
We did not include correlated noise between pixels, although this does happen for real images in the HST drizzling pipeline.

The U-Net was trained to classify each pixel within the image.
The pixels where the profile mass density of the main lens is larger than the critical density (the convergence is greater than 1) are labeled as ``main lens.''
We defined pixels for the subhalos by drawing a circle around their location with a radius of 2 pixels.
The value of the label was chosen by mass bins logarithmically spaced as $\{10^{6},$ $10^{6.5}$, $10^{7}$, $10^{7.5}$, $10^{8}$, $10^{8.5}$, $10^{9}$, $10^{9.5}$, $10^{10}\}\msun$.
We refer to pixels that are not labeled as the main lens or as subhalos as ``background.''

Our U-Net was trained on images with either no substructure or exactly one subhalo.
The training set comprised $9\times10^6$ images, with one-tenth having only a smooth lens and the other nine-tenths additionally containing one subhalo that falls into one of nine mass bins.
The task of the network is to predict the class for each pixel within the image from one of 11 categories (the main lens, a subhalo within one of the nine mass bins, or neither).
While at face value this is a classification problem, we note that it simultaneously locates and estimates the mass of the subhalos.
Further details of the U-Net architecture and the training can be found in the appendix.

In this work we train networks on four different data sets, a simple lens and a complex lens with either one or 50 orbits of exposure.

\section{Reconstructing the lens and resolving substructure}

\begin{figure}[t]
    \centering
    \includegraphics[width=\linewidth]{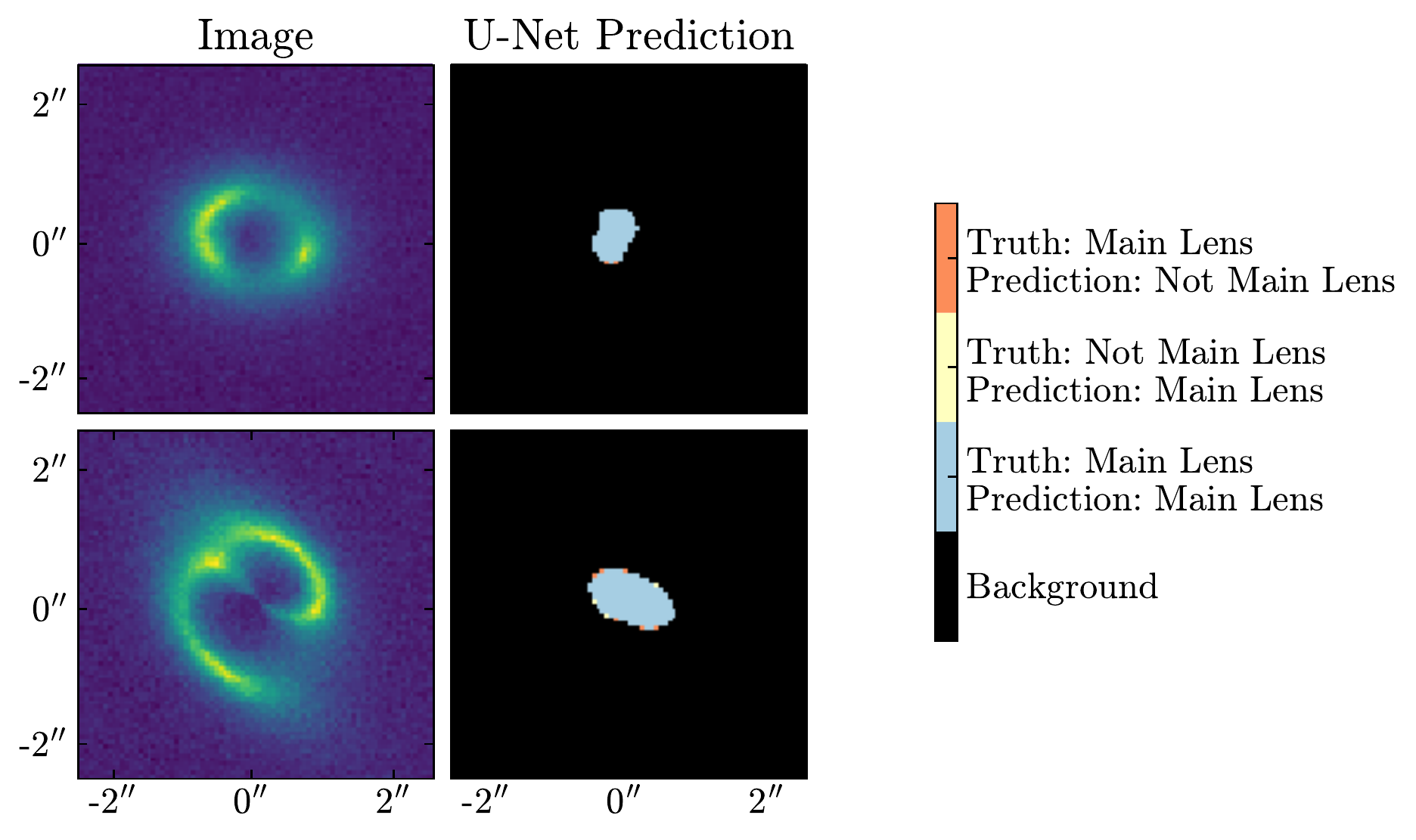}
    \caption{Reconstruction of the main lens. The left panels show the input strong lensing images, which contain a complex lens without substructure and with one orbit of simulated HST exposure as well as a source with an apparent magnitude of 22. The right panels show the output of the neural network.
    }
    \label{fig:smooth_lens_only}
\end{figure}

The networks were tested against an independent set of $5\times 10^5$ images, which were drawn from the same population as the training set.
First, we examined the images that only contain a smooth lens and no substructure.
Figure~\ref{fig:smooth_lens_only} shows two example images coming from a complex lens with 1 orbit of exposure time.
The left panels show the input images, with HST-like noise, and the right panels show the network predictions.
We have colored the pixels according to the network accuracy: blue is correct, orange pixels were supposed to be part of the main lens but were missed (false negatives), and yellow were not supposed to be part of the main lens but are predicted to be so by the network (false positives).
In the upper panels the lens has substantial multipole moments, and in the lower panel the lens is close to elliptical.
In both cases, the network correctly predicts the shape and size of the lens to within a few percent.
Over the entire test sample with only a smooth lens, $98.7\%\pm2.0\%$ of the pixels that should be labeled as the main lens are labeled correctly.
Only $1.3\%\pm2.0\%$ of the area of the true main lens contains pixels that were supposed to be labeled as the main lens but were missed.
Another $1.2\%\pm1.8\%$ of the area of the true main lens is incorrectly added to the shape of the lens.
The fact that the false positives and false negatives account for the same area of the main lens leads to the area of the predicted subhalo being $99.9\% \pm 1.6\%$ of the true area across the whole testing set.
Image segmentation can accurately extract complex lenses, which could be useful for applications that need carefully modeled lenses, for instance when extracting the value of the Hubble constant with lensed quasars~\citep{2020A&A...639A.101M}.

\begin{figure}[t]
\centering
\includegraphics[width=\linewidth]{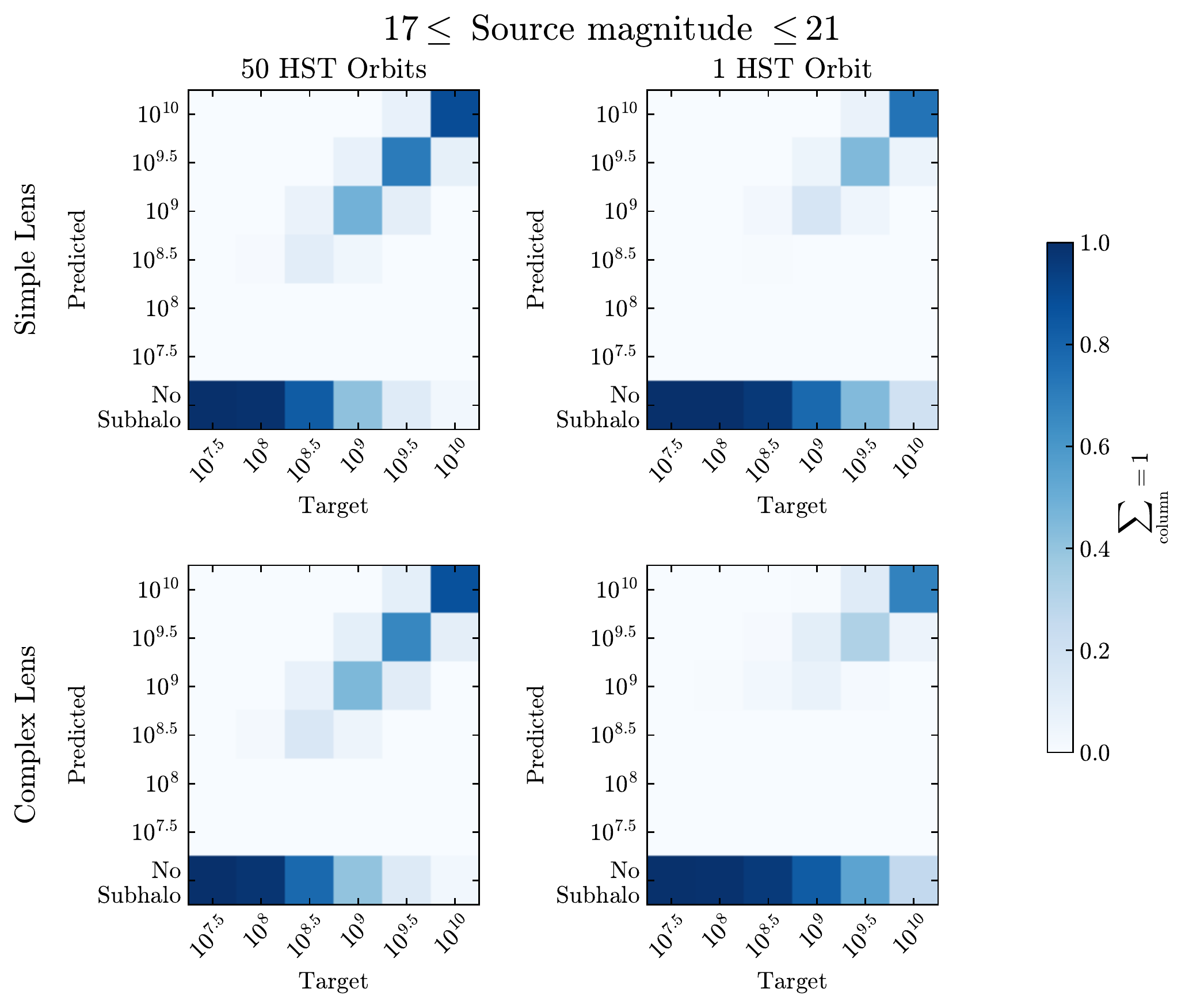}
\caption{{Confusion matrix for subhalos in the test images.
The matrix is normalized such that the sum of the columns is unity. 
}
}
\label{fig:CM}
\end{figure}

Now, we shift our attention to the substructure in the lens.
In Fig.~\ref{fig:CM} we show the confusion matrix for the test set.
The columns correspond to the true class (target) of a subhalo, while the rows show what the model classifies it as (predicted).
We normalized the columns such that they sum to unity: with this choice, the values in each column show the fraction of subhalos with a given true label that are predicted to be in each of the possible classes.
The left panels correspond to models trained and tested on images with 50 orbits of HST exposure, while the right panels correspond to models trained and tested on images with one orbit.
The upper and lower panels correspond to the simple lens and the complex lens, respectively.
In each panel, the matrix is mostly diagonal, implying remarkable accuracy overall.
However, the network struggles to detect the lightest subhalos.

\renewcommand{\arraystretch}{1.3}
\setlength{\tabcolsep}{4 pt}
\begin{table*}[t]
    \centering
    \caption{\small{
    Percentage of subhalos detected (predicted to be in the correct or adjacent mass bin) across the test set with apparent source magnitudes between [17-21].}}
    \begin{tabular}{l| c c c c c}
    \hline
    \hline
     & $10^{8}\msun$ & $10^{8.5}\msun$ & $10^{9}\msun$ & $10^{9.5}\msun$ & $10^{10}\msun$\\
     \hline
    50 HST Orbits, simple lens& 0.8 & 17.1 & 58.9 & 87.9 & 97.1 \\
    1 HST Orbit, simple lens& 0.0 & 3.3 & 22.1 & 55.5 & 80.2 \\
    \hline 
    50 HST Orbits, complicated lens& 2.0 & 22.1 & 59.3 & 86.7 & 96.9 \\
    1 HST Orbit, complicated lens& 0 & 2.8 & 16.6 & 45.3 & 73.9 \\
    \hline
    \hline
    \end{tabular}
    \label{tab:Accuracies}
\end{table*}

We see that there is always a nonzero probability of subhalos getting assigned to the adjacent mass bins by the model.
When the network misclassifies a subhalo, it is often still locating a subhalo but getting a slightly higher or lower mass estimate. 
This is not surprising since the subhalo masses can lie near the boundaries of adjacent classes.

Comparing the different exposures allows us to examine the effects of both noise and the complexity of the system in detecting subhalos with image segmentation.
For instance, with 50 orbits in the simple system, the network is able to identify 17\% of the subhalos in the $10^{8.5}\msun$ bin and correctly tag more than 87\% of the subhalos with $m\geq10^{9.5}\msun$.
Complexity in the lens does not lead to a decrease in performance -- 86\% of the subhalos in the $10^{9.5}\msun$ bin are still accurately predicted, and the accuracy of the $10^{8.5}\msun$ pixels changes to 22\%.
We note that all of these percentages are derived from subhalos that are in pixels at least half as bright as the brightest pixel, and the accuracy would go up or down if we examined a brighter or dimmer cut.
Similarly, we used bright sources (between magnitude 17-21), and the accuracy goes down for dimmer sources.

When the exposure time is decreased to a single orbit, we see that the accuracy for subhalos with $m\leq10^{8.5}\msun$ is less than 5\%.
Pixels from subhalos in the $10^{9.5}\msun$ bin are correctly predicted 55\% and 45\% of the time for the simple and complex systems, respectively.
In Table~\ref{tab:Accuracies} we show a summary of the confusion matrix by computing the percentage of the subhalos detected by the network and placed in either the correct or adjacent mass bin.

\section{Domain adaptation}
\label{sec:Generalization}

It is challenging to get ML models to work when new data are outside the realm of what they have been trained on (see \citealt{2017arXiv170205374C} and \citealt{WANG2018135} for reviews).
Both in the training and in the testing presented, each image had either no substructure or exactly one subhalo.
Due to the steep slope of the subhalo mass function, we expect that many subhalos should be present in strong lensing images~\citep{2008MNRAS.391.1685S}.
We therefore assessed whether the network is capable of generalizing to other lensing situations after training.

\begin{figure}[t]
\centering
\includegraphics[width=\linewidth]{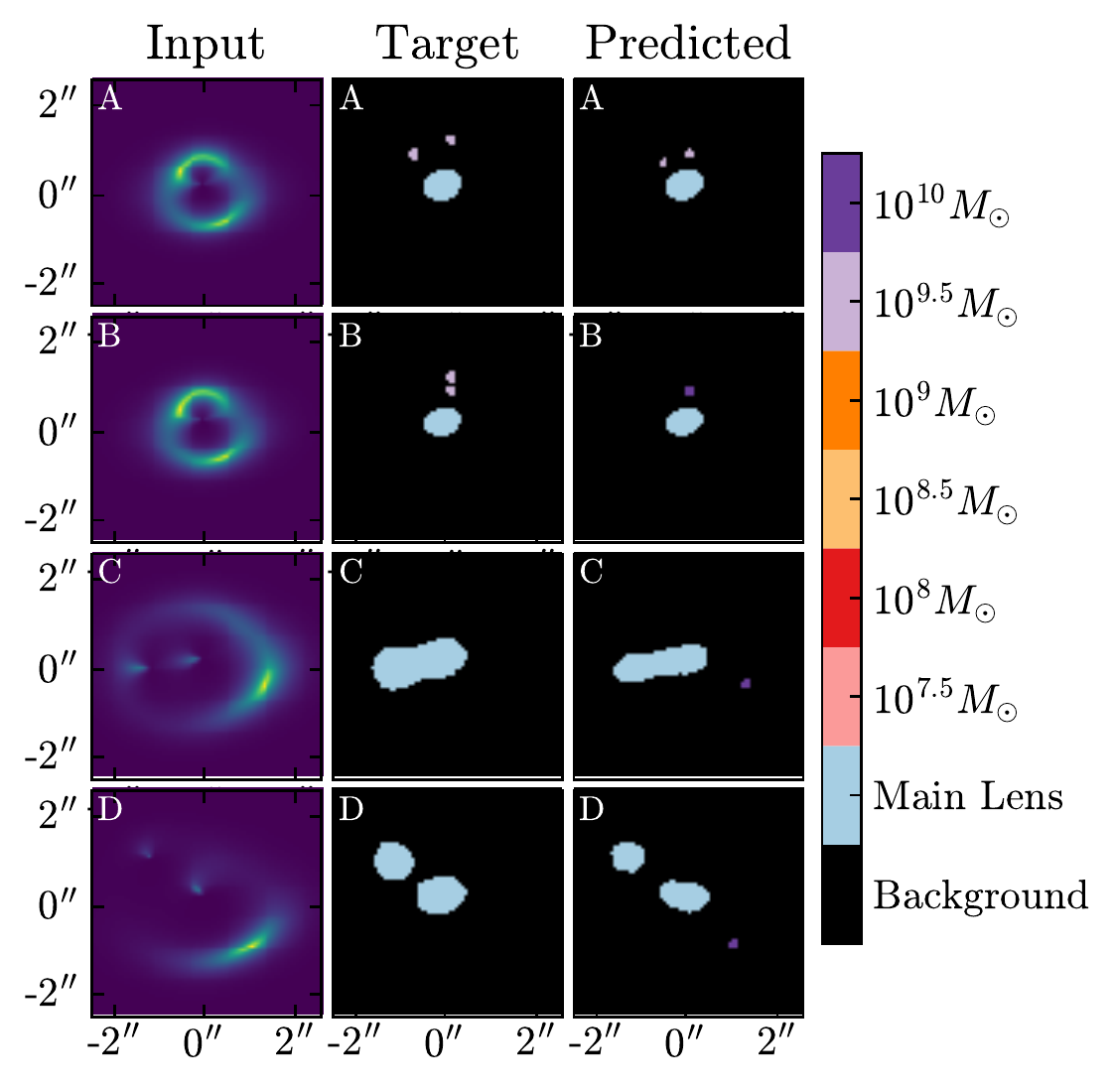}
\caption{Results when the network is applied outside of the domain of the training data. The left column displays the strongly lensed images used as input to the neural network.
The middle column contains the pixel-by-pixel labels we want the network to reproduce.
The right column shows the network predictions.
The input images assume one full orbit of exposure for an apparent magnitude-19 source.
}
\label{fig:SpecialCases}
\end{figure}

In Fig.~\ref{fig:SpecialCases} we examine the effect of two subhalos being close or overlapping.
We chose two subhalos in the $10^{9.5}\msun$ bin.
The main lens is an SIE with an Einstein radius of $1\arcsec$ and an ellipticity of $q=0.75$; the source and main lens are kept constant throughout these images to only see the effect due to subhalos.
In panel A, the two subhalos are far enough away from each other that the network is able to resolve them separately.
In panel B, the two subhalos are close enough that the true pixels are nearly touching each other.
The network does not correctly identify two individual subhalos in this case, but it does classify the pixels as belonging to a single, higher-mass subhalo of $10^{10} M_{\odot}$.
An animation of the subhalo traversing the image and its effect on the network output can be found at \href{https://bostdiek.github.io/Videos/Overlapping_v4.mp4}{this link}.\footnote{\url{https://bostdiek.github.io/Videos/Overlapping_v4.mp4}}

Finally, in panels C and D of Fig.~\ref{fig:SpecialCases}, a second SIE halo is included in the image with an Einstein radius of $0.8\arcsec$ and no ellipticity.
The combined lenses result in much larger distortions of the light than seen during training, as shown in the input images in the left columns.
While the network is capable of identifying the presence and location of a second main lens, its shape is not well captured.
In fact, we can see in panel D how it adds a spurious heavy subhalo to account for the extra distortion.
An animation showing the different positions of the second lens, and its effect on pixel classification, is available at \href{https://bostdiek.github.io/Videos/TwoMain_v3.mp4}{this link}.\footnote{\url{https://bostdiek.github.io/Videos/TwoMain_v3.mp4}} 

\section{Discussion and outlook}

This Letter presents a new method for analyzing images of strong gravitational lensing.
We show that image segmentation using a U-Net is able to resolve the main lens, even when it has large multipole moments.
The network is also able to determine both the mass and location of dark matter substructure.
For subhalos located in bright pixels near the Einstein ring, the U-Net detects 18.1\% of the $10^{9}\msun$ subhalos and 61.9\% of the $10^{9.5}\msun$ subhalos.

While the network was trained only on images with a single strong lens and low-mass perturber near the Einstein ring, it is able to detect the mass and location of additional substructure. 
When subhalos overlap, the network labels the sum of their effects.\footnote{
We note, however, that our method cannot obtain covariances between models with differing numbers of subhalos, as in \cite{2018ApJ...854..141D}}
Furthermore, we have shown that the network is also capable of identifying two separate main lenses.
These results suggest that the network has learned a powerful representation of the physics involved in our lensing setup. 
Domain adaptation is a notoriously difficult task for ML models, but the network is able to generalize very well to images that look very different from the training samples but are nevertheless governed by the same principles. 

The success of our image segmentation technique is encouraging for various science applications of strong lens images. 
It paves the way for a new method of extracting the subhalo mass function from strong lens images, which we do for simulated images in \citep{Ostdiek:2020mvo}. 
The subhalo mass function is a key target for dark matter science as we can use it to diagnose deviations from the CDM scenario.
Additionally, the ability to characterize non-elliptical lenses could be useful for other lensing applications.

In sum, when including realistic levels of noise, we reach good accuracy for the same range of masses detected by traditional methods, but while these take several weeks to analyze each system, we do it in a fraction of a second.
Furthermore, the success of the network in generalizing to multiple main lenses can also be seen as a demonstration of the network's knowledge of the main lens.

Image segmentation is a promising new technique for studying strong lensing systems.
However, there are still steps to solve before applying it to real data.
For instance, both our lenses and sources were at fixed redshift.
We have not characterized how the network accuracy would change if these assumptions were removed.
Since we also assumed that the light from the lens could be perfectly removed, further studies need to be performed for applying image segmentation to images with imperfect lens light subtraction.
Finally, complications in the source light can mimic the effects of subhalos.
Further study is necessary to improve the robustness to more realistic sources.

\begin{acknowledgements}
We thank Simon Birrer, A. Cagan Sengul, and Arthur Tsang for help with the code \textsc{Lenstronomy}.
We also thank Simon Birrer, Johann Brehmer, Tim Cohen, and Siddharth Mishra-Sharma for helpful comments on a previous version of this manuscript.
BO was supported in part by the U.S. Department of Energy under contract DE-SC0013607.
CD was partially supported by the Department of Energy (DOE) Grant No. DE-SC0020223.
The computations in this paper were run on the FASRC Cannon cluster supported by the FAS Division of Science Research Computing Group at Harvard University.
\end{acknowledgements}


\bibliographystyle{aa}
\bibliography{segmentation}

\clearpage
\appendix

\renewcommand{\theequation}{A.\arabic{equation}}
\renewcommand{\thefigure}{A.\arabic{figure}}
\renewcommand{\thetable}{A.\arabic{table}}

\section{Data generation}
\label{sec:gen}

In the main text we discuss the general framework used for defining our simple and complex lenses.
In this section we provide more details for image generation using the \texttt{Lenstronomy} software package~\citep{2018PDU....22..189B, 2015ApJ...813..102B}.
Each image contains source light, a smooth lens model, and sometimes substructure within the lens.
We studied images that are $80\times 80$ pixels and cover an area of $5\arcsec \times 5\arcsec$, giving a resolution of $0.06\arcsec$.
Distances were computed using the \emph{Planck} 2015 cosmology~\citep{Ade:2015xua}.

We treated the \textbf{source light} as a galaxy at redshift $z_{\rm{source}} = 0.6$.
The light source always has a Sersic profile.
The Sersic index controls the degree of curvature of the profile with radius and was chosen to be in the range $n_{\rm{ser},i} \in [0.5, 1].$
The angle of the major axis was chosen randomly in the range $-\pi$ to $\pi$.
The location of the source was near the center of the image, with the $x$ and $y$ coordinates drawn from a multivariate Gaussian with zero mean and diagonal covariance matrix with $\sigma_{xx}^2=\sigma_{yy}^2=0.001$.
The radius of the light source was in the range [0.1-0.8] kpc.
The apparent magnitude of the source was chosen uniformly in the range [17-25].
This range of magnitudes created a wide domain of signal-to-noise ratios for the training data.

The \textbf{smooth lens} was modeled with an SIE~\citep{1994A&A...284..285K} at redshift $z_{\rm{lens}} = 0.2$.
This profile was parameterized by the Einstein radius and the ellipticity.
The Einstein radius was chosen to be between $\theta_E \in [0.85, 1.50] \arcsec$, and the ellipticity was chosen between $q \in [0.4, 1]$.
The coordinates of the lens were chosen uniformly as $x,y \in [-0.25, 0.25]\arcsec$.
To allow for more realistic lenses, we additionally added $m=3$ and $m=4$ multipole moments for the lensing potential to capture departures from ellipticity.
The amplitude of each moment was chosen uniformly between $[0, 0.1],$ and the orientation could take on any angle.
External shear was also added to the lens, drawn uniformly between $[-0.2, 0.2]$ in the $x$ and $y$ directions.

\textbf{Substructure} was added to the lens, modeled as truncated NFW profile subhalos~\citep{1996ApJ...462..563N,2009JCAP...01..015B}.
In this work we used a fixed concentration for the profiles with $c=15$.
We chose the truncation radius as five times the scale radius.
For both training and testing the network, we only placed the subhalos in pixels that are at least 50\% as bright as the brightest pixel in an image.

With the lens defined, we were able to generate the target labels that the network is trying to predict.
First, we examined the lens without substructure.
Any pixels that have a convergence greater than 1 are defined to belong to the main lens class.
Then we marked a circle with a radius of two pixels at the location of the subhalos and labeled those pixels with which mass bin the subhalo falls into.
The remaining pixels that were not labeled as the main lens or the subhalos are called background pixels.

We simulated \textbf{noise and detector effects} using the \texttt{SimulationAPI} within \texttt{Lenstronomy}.
We selected the WFC3\_F160W camera band and used the built in pixel-based PSF.
The exposure time was set to 5400~s for 1 orbit or 270000~s for 50 orbits.
The sky brightness was magnitude 22.3 and the zero point magnitude was 25.96.
Additionally, we set the supersampling to 1.
\section{Network architecture and training}
\label{sec:train}

To perform the image segmentation task, we implemented a U-Net model~\citep{2015arXiv150504597R} with \textsc{PyTorch}~\citep{NEURIPS2019_9015}.
The input to the network was an $80\times80$ pixel image with a single channel (a grayscale image).
We normalized the image by dividing by the maximum value.
The U-Net consists of two halves: a contracting path (where the image dimension shrinks) that is followed by an expanding path (where the image dimension is dilated again).

The U-Net was built by repeating a series of three operations.
These operations are convolution, batch normalization, and rectified linear unit (ReLU) activation, which were then repeated a second time.
We refer to this combination of operations as a ``block.''
During the first block, the convolutional layers have a height and width of $3\times3$ and 64 filters (resulting in 64 channels).
The stride and padding on the convolution block were set to keep the height and width unchanged throughout the block while allowing for more channels.

The resulting image was then down-sampled to $40\times40$ pixels using a max pooling layer.
The down-sampling allows the next layers to effectively apply to a larger area of the image, giving the network the ability to learn features on different scales.
From here, another block with 128 filters was applied.
The result was again down-sampled to $20\times20$ pixels.
Next was a block with 256 filters.
The final contraction was down-sampled to $10\times10$ pixels.
Additional blocks were used, with filter sizes of 512 and 256 in the two convolutional layers, respectively.

In the expansion path, lower resolution information was up-sampled to higher resolution.
This was done with a large number of feature channels, which allows the network to propagate information to the high resolution layers.
The up-samplings were done with a transposed convolutional layer with a height and width of $3\times3$ in \textsc{PyTorch}.

The output of the last layer in the contracting path is a $10\times10\times 256$ array.
In the first step of the expanding path the first two dimensions were up-sampled to $20\times20$ and then concatenated with the last layer in the contracting path with the same resolution.
Following the concatenation, the data have dimensions of $20\times20\times512$.
This was followed by a block with 256 and 128 filters.
The data were again up-sampled, now to $40\times40,$ and concatenated with the corresponding layer of the contracting path.
A block was again applied, with 128 and 64 filters.

A final up-sampling brought the data back to the original resolution.
This was then concatenated with the last layer before down-sampling.
A block was applied, with 64 and 11 filters.
These final 11 filters correspond to the 11 different classes the pixels can belong to (main lens, one of nine subhalo mass bins, or background).
The softmax activation function was applied along the channel dimension such that the sum of the 11 features for each pixel is 1.
Thus, the pixel channels correspond to the probability of a pixel belonging to each class.

The cross entropy loss function was computed for each pixel, and the network parameters were tuned using the Adam optimizer~\citep{2014arXiv1412.6980K} with the standard $\beta$ values to minimize the loss.
We started with a learning rate of $10^{-3}$ and iterated through the data in batches of 100 images.
The loss was evaluated on an independent validation set after each epoch.
When the validation loss did not improve for five epochs, the learning rate was lowered by a factor of ten; we did not allow the learning rate to drop below $10^{-6}$.
The training ended when the validation loss did not improve for 15 epochs.

\section{Smooth lens finding}
\label{sec:mainlens}

\begin{figure}[h]
    \centering
    \includegraphics[height=7.in]{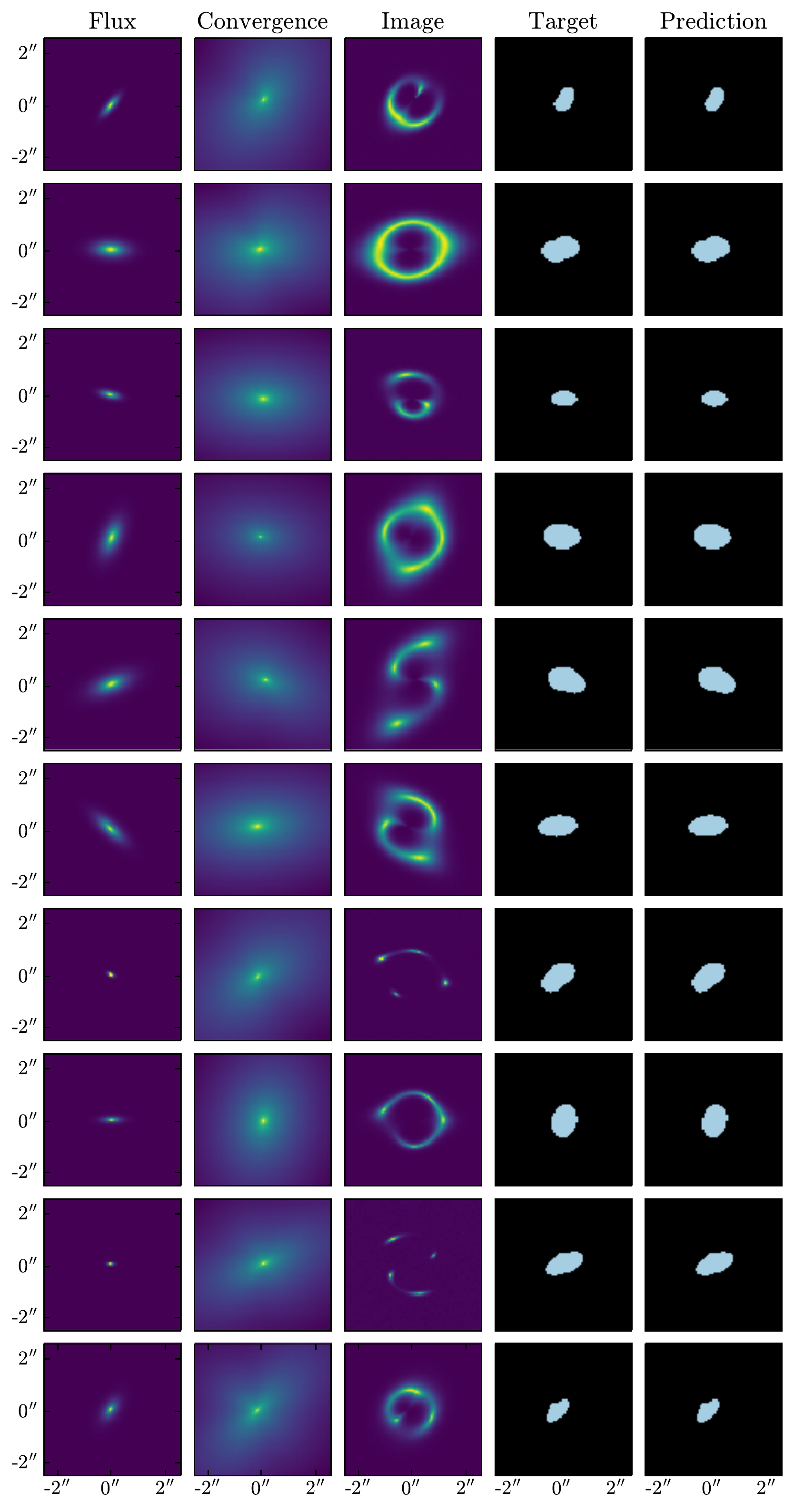}
    \caption{Examples of the U-Net being applied to lenses with no substructure.
    In the target and prediction columns, the blue pixels indicate the main lens (or areas where the convergence is above unity).
    The network is able to accurately describe the main lens for a wide variety of lens shapes and sizes.}
    \label{fig:example_main_lens}
\end{figure}

In Fig.~\ref{fig:example_main_lens} we show many examples of the U-Net classifying images without substructure. 
The left column shows the flux from the source light (if no lens was present). 
The second column shows the convergence field on a logarithmic color scale. 
The third column shows the image that is input to the network and contains noise from exposure of one HST orbit.
The fourth column shows which pixels have a convergence greater than 1 at truth-level and are marked as belonging to the main lens.
The final column contains the predictions from the U-Net.
The network is accurate for both the larger and smaller lenses as well as for elliptical lenses and lenses with departures from ellipticity.

\end{document}